%% file: aams.tex
\documentclass[british,nofootinbib,aps,prd,a4paper,T1,hyphens,spaces,reprint]{revtex4-2}%preprint: easer to proofread, reprint: closer to journal typesetting
\usepackage[T1]{fontenc}%Umlaute und saubere Worttrennung bei solchen Wörtern
\usepackage{microtype}%stellt "schöne" Worttrennungen und Zeilenumbrüche sicher
\usepackage[utf8]{inputenc}
\usepackage[british]{babel}%Schreibweisen und ""'' und so Sachen
\usepackage{setspace}%Abstände
\usepackage{enumerate}
\usepackage{csquotes}%sets correct quotation marks
\usepackage{color}

\usepackage{float}
\usepackage{graphicx}
\usepackage{tabularx}
\usepackage[inline, final]{showlabels}
\usepackage{booktabs}
\usepackage[title,titletoc]{appendix}%Allows for appendices
\usepackage[caption=false]{subfig}
\usepackage{flafter} % avoid figures from displaying before they were defined
\usepackage{placeins}%Provides FloatBarrier

\usepackage{amsmath}%Mathe-Notationen

\usepackage{amssymb}% Für mathematica code direkt einzubinden
	\newcommand{\mathsym}[1]{{}}
	\newcommand{\unicode}[1]{{}}
\usepackage{amstext}%Falls man Text in einer Gleichung möchte
\usepackage{tensor}
\usepackage[separate-uncertainty=true,multi-part-units=single]{siunitx}
\usepackage{physics}
\usepackage[nonumberlist,nopostdot,numberedsection=autolabel]{glossaries}
\setacronymstyle{long-short}
\loadglsentries{abbrv}

\usepackage{varioref}%Verweise
\PassOptionsToPackage{hyphens}{url}%sollte Probleme mit zu langen URLs lösen
\usepackage[pdfencoding=auto,psdextra,unicode=true,pdfusetitle,bookmarks=true,bookmarksnumbered=false,bookmarksopen=false, breaklinks=true,pdfborder={0 0 0},backref=false,colorlinks=false] {hyperref}%funktionierende Links in PDFs
\usepackage{cleveref}%Macht z.B. nicht nur "(1)" sonder gleich "Gleichung (1)"; nutzen mit \cref

\definecolor{webgreen}{rgb}{0, 0.5, 0} % less intense green
\definecolor{webblue}{rgb}{0, 0, 0.5} % less intense blue
\definecolor{webred}{rgb}{0.5, 0, 0} % less intense red

\makeatletter
\g@addto@macro\bfseries{\boldmath}  % automatic bold math in sections
\makeatother

\begin{document}
\title{AI Insights into Theoretical Physics and the Swampland Program: A Journey Through the Cosmos with ChatGPT}
\author{Kay Lehnert}
\email{k.lehnert@protonmail.com}
\affiliation{Department of Theoretical Physics, Maynooth University, Maynooth, Ireland}

\date{\today}

\begin{abstract}
	\input{S0_Abstract}
\end{abstract}

\maketitle

\tableofcontents
\input{S1_Text.tex}

\bibliography{references,bib,MA}

\end{document}

%% file: abbrv.tex
%\newacronym[longplural={interstellar media}]{ism}{ISM}{interstellar medium}
\newacronym{rhs}{RHS}{Right-Hand Side}
\newacronym{lhs}{LHS}{Left-Hand Side}
\newacronym{cmb}{CMB}{Cosmic Microwave Background}
\newacronym{wgc}{WGC}{Weak Gravity Conjecture}
\newacronym{adscft}{AdS/CFT}{Anti-de~Sitter/Conformal Field Theory}
\newacronym[\glsshortpluralkey={EoS},longplural={Equations of State}]{eos}{EoS}{Equation of State}
\newacronym[\glsshortpluralkey={EoM},longplural={Equations of Motion}]{eom}{EoM}{Equation of Motion}
\newacronym{bao}{BAO}{Baryon Acoustic Oscillations}
\newacronym{flrw}{FLRW}{Friedmann–Lemaître–Robertson–Walker}
\newacronym[name={$\bm{\Lambda}$CDM}]{lcdm}{$\Lambda$CDM}{$\Lambda$ Cold Dark Matter}
\newacronym{geodes}{GEODEs}{GEneric Objects of Dark Energy}
\newacronym{svt}{SVT}{Scalar-Vector-Tensor}
\newacronym{de}{DE}{Dark Energy}
\newacronym{dm}{DM}{Dark Matter}
\newacronym{tcc}{TCC}{Trans-Planckian Censorship Conjecture}
\newacronym{trgb}{TRGB}{Tip of the Red Giant Branch}
\newacronym{bbn}{BBN}{Big Bang Nucleosynthesis}
\newacronym{aic}{AIC}{Akaike Information Criterion}
\newacronym{bic}{BIC}{Bayesian Information Criterion}
\newacronym{kids}{KiDS}{Kilo Degree Survey}
\newacronym{boss}{BOSS}{Baryon Oscillation Spectroscopic Survey}
\newacronym{2dflens}{2dFLenS}{2-degree Field Lensing Survey}
\newacronym{des}{DES}{Dark Energy Survey}
\newacronym{viking}{VIKING}{Visible and Infrared Survey Telescope for Astronomy (VISTA) Kilo-Degree Infrared Galaxy Survey}
\newacronym{bps}{BPS}{Bogomol'nyi–Prasad–Sommerfield}
\newacronym{aligo}{aLIGO}{advanced Laser Interferometer Gravitational-Wave Observatory}
\newacronym{lisa}{LISA}{Laser Interferometer Space Antenna}
\newacronym{shoes}{SH0ES}{Supernovae and $H_0$ for the Dark Energy Equation of State}
\newacronym{eboss}{eBOSS}{extended Baryon Oscillation Spectroscopic Survey}
\newacronym{holicow}{H0LiCOW}{$H_0$ Lenses in COSmological MOnitoring of GRAvItational Lenses (COSMOGRAIL)'s Wellspring}
\newacronym[\glsshortpluralkey={SNe},longplural={supernovae}]{sn}{SN}{supernova}
\newacronym{calpriorsnia}{CalPriorSNIa}{effective Calibration Prior on the absolute magnitude of Type Ia Supernovae}
\newacronym[\glsshortpluralkey={QFTs},longplural={Quantum Field Theories}]{qft}{QFT}{Quantum Field Theory}
\newacronym{gr}{GR}{General Relativity}
\newacronym{qg}{QG}{Quantum Gravity}
\newacronym[\glsshortpluralkey={EFTs},longplural={Effective Field Theories}]{eft}{EFT}{Effective Field Theory}

%% file: S0_Abstract.tex
%\section*{Abstract} \label{s:abstract}
In this case study, we explore the capabilities and limitations of ChatGPT, a natural language processing model developed by OpenAI. We find that it is effective at paraphrasing and explaining concepts in a variety of styles, but not at genuinely connecting concepts. It will provide false information with full confidence and make up statements when necessary. However, its ingenious use of language can be fruitful for identifying analogies and describing visual representations of abstract concepts.

%\vfill
Keywords: --- \texttt{Cosmology} --- \texttt{Swampland Programme} --- \texttt{String Theory} --- \texttt{String-Theoretical Conjectures} --- \texttt{Swampland Conjecture} --- \texttt{de~Sitter Conjecture} --- \texttt{Weak Gravity Conjecture} --- \texttt{ChatGPT} --- \texttt{Artificial Intelligence} --- \texttt{Natural Language Processing Model} ---

%% file: S1_Text.tex
\section{Introduction}
\texttt{ChatGPT}\footnote{\url{https://chat.openai.com/chat}} is a \textit{chatbot} by OpenAI launched on 30 November \cite{openai_chatgpt_2022}. Five days later, it had over 1 million users \cite{sam_altman_sama_chatgpt_2022}. It has since gained a lot of traction---social media is flooded with excerpts of conversations with \texttt{ChatGPT} that are often surprisingly human-like, funny, and insightful. Some scholars are concerned that the end of essays as a meaningful assessment tool might be near \cite{stokel-walker_ai_2022,yeadon_death_2022}, since \texttt{ChatGPT} can easily generate convincing essays, paragraphs, or chapters for any given topic. In this letter, we explore how well it copes with questions regarding the rather niche topic of string theoretical swampland conjectures. Swampland conjectures are an intriguing application of \texttt{ChatGPT}, because they synthesise different branches of theoretical physics: general relativity and quantum field theory in the framework of sting theory, applied to cosmology. Despite the fact that they contain a heavy mathematical formalism, a lot of the work is still done on a conceptual level. This favours \texttt{ChatGPT}'s strong language capabilities. We examine how strong \texttt{ChatGPT}'s skills are, when it comes to combining, mixing, and synthesising the relevant concepts. Furthermore, we explore how we can use \texttt{ChatGPT} to our advantage in research, outreach, and education.

This letter presents excerpts from a conversation\footnote{The full conversation can be found on \url{https://github.com/kabeleh/ChatGPT}.} between the author and \texttt{ChatGPT}, which begins with broader and more general questions about the field of string theory, before narrowing it down to specific swampland conjectures\footnote{See e.g. \citet{vafa_cosmic_2019,palti_swampland_2019} for an overview.} and \texttt{ChatGPT}'s \textit{understanding} of them. We attempt to highlight strengths and weaknesses of \texttt{ChatGPT}, present some use cases, and raise some red-flags.

We start with the question \enquote{Can you give me a definition of ChatGPT?}.\footnote{Throughout the text, \texttt{ChatGPT}'s responses are written in typewriter. \texttt{ChatGPT} is used in its December 15 version.}
\begin{displayquote}
    \texttt{ChatGPT is a natural language processing (NLP) model developed by OpenAI that is specifically designed to generate human-like text in real-time chat environments. It is based on the GPT (Generative Pre-training Transformer) model, which is a type of machine learning model that has been trained on a large dataset of human-generated text and is capable of generating text that is similar to human writing in terms of style, structure, and content.}

    \texttt{The purpose of ChatGPT is to enable the creation of chatbots or conversational agents that can engage in natural and engaging conversations with users. ChatGPT can be used in a variety of applications, such as customer service chatbots, personal assistants, or educational chatbots. It can generate responses to user input based on the context of the conversation, allowing it to carry on a conversation in a way that is natural and human-like.}
\end{displayquote}

 \texttt{ChatGPT} has a vast knowledge of a variety of subjects and can produce surprisingly elegant or creative explanations of concepts. It can be used to obtain definitions and explanations of concepts in different styles, e.g. to explain string theory to a high-school student: \begin{displayquote}
    \texttt{String theory is a theoretical framework in physics that attempts to reconcile quantum mechanics and general relativity, two of the fundamental theories that describe the behavior of the universe. It is based on the idea that the fundamental building blocks of the universe are not particles, but rather tiny, vibrating strings.}

    \texttt{According to string theory, the fundamental forces of nature (such as gravity, electromagnetism, and the strong and weak nuclear forces) are not caused by the exchange of particles, but rather by the vibration of these strings. Different patterns of vibration correspond to different particles and forces.}

    \texttt{One of the key predictions of string theory is the existence of extra dimensions beyond the three dimensions of space and one dimension of time that we are familiar with. These extra dimensions may be hidden from our everyday experience, but they could potentially play a role in the behavior of particles and forces at very small scales.}

    \texttt{Although string theory is still being developed and is not yet a fully developed theory, it has the potential to provide a deeper understanding of the fundamental nature of the universe and could potentially resolve some of the outstanding problems in physics.}
\end{displayquote}

Using \texttt{ChatGPT}'s ability to impersonate a predefined persona can help the user to find the right words for her target audience. \texttt{ChatGPT} can be asked to rephrase a given paragraph in a given style, or can directly be asked to present the answer in the words of a character. The given answer regarding string theory covers all relevant points and is adequate for a high-school student. Furthermore, if there is no simple answer or the answer is less clear, an attempt will be made to provide one. As the following response demonstrates, \texttt{ChatGPT}'s response to the question \enquote{What is the most important equation in string theory?} is sound and even includes equations that frequently appear in string theory.
\begin{displayquote}
    \texttt{It is difficult to identify a single 'most important' equation in string theory, as the theory is still being developed and is not yet a fully formulated and complete theory. String theory is a theoretical framework that attempts to reconcile quantum mechanics and general relativity, and it involves a wide range of mathematical concepts and equations.}

    \texttt{One equation that is often associated with string theory is the Euler-Lagrange equation, which is a mathematical expression that describes the motion of a system in terms of its energy. This equation plays a role in the formulation of string theory, as it is used to describe the behavior of strings and the forces that act on them.}

    \texttt{Another important equation in string theory is the Schrödinger equation, which is a mathematical expression that describes how the quantum state of a system changes over time. The Schrödinger equation is a fundamental equation of quantum mechanics and is used to describe the behavior of particles on the quantum level.}

    \texttt{It is worth noting that string theory is a very complex and interdisciplinary field, and there are many other equations and mathematical concepts that are important to the development of the theory. These include the Einstein field equations, which describe the fundamental nature of gravity, and various equations that describe the behavior of particles and forces at very small scales.}
\end{displayquote}

As a follow-up question, we asked \texttt{ChatGPT} what those equations look like. The answer, along with general explanations of what the different terms mean and what the equations describe, was:
\begin{align}
    L - \frac{\partial L}{\partial q}\frac{\partial q}{\partial t} &= \frac{\partial L}{\partial q^\prime}\frac{\partial q^\prime}{\partial t}\\
    i\hbar \frac{\partial \Psi}{\partial t} &= H\Psi\\
    G_{\mu\nu} &= 8\pi T_{\mu\nu}\\
    &=R_{\mu\nu} - (1/2)g_{\mu\nu}R + \Lambda g_{\mu\nu},
\end{align}
where the expansion of the Einstein equation in terms of the Ricci tensor and Ricci scalar was only provided on explicit request.

\section{Swampland Conjectures}
To put \texttt{ChatGPT}'s knowledge of swampland conjectures to the test, we asked it \enquote{What do you know about the string theoretical swampland program?}:
\begin{displayquote}
    \texttt{The swampland program is a research program in string theory that aims to identify and study the constraints on the possible low-energy effective theories that can emerge from string theory. It is based on the idea that string theory may contain a vast landscape of possible theories, and that only a small subset of these theories correspond to physically realistic models of the universe.}

    \texttt{The swampland program seeks to identify the criteria that must be satisfied by a low-energy effective theory in order for it to be consistent with string theory. These criteria are known as the swampland conditions. The swampland program has led to the development of several important conjectures and criteria, including the swampland distance conjecture, the swampland de Sitter conjecture, and the swampland conjectures on trans-Planckian censorship.}

    \texttt{The swampland program is a highly active and influential research program in string theory and has attracted a great deal of attention and interest in the scientific community. It has the potential to provide important insights into the fundamental nature of the universe and the structure of string theory.}
\end{displayquote}

It provides three examples of swampland conjectures in its response: the distance conjecture \cite{palti_swampland_2019,Ooguri_2007,Ooguri_2019,grana2021swampland,vafa_cosmic_2019}, the de~Sitter conjecture \cite{vafa_cosmic_2019,obied2018sitter,palti_swampland_2019,Denef_2018,Ooguri_2019,Agrawal_2018,andriot2019refining} as well as the trans-Planckian censorship conjecture \cite{Bedroya_2020,Bedroya_2020b,palti_swampland_2019}. There are other conjectures. \texttt{ChatGPT} gladly provides a list of 25 conjectures upon request:
\begin{displayquote}
    \begin{enumerate}
        \item\texttt{The distance conjecture}
\item\texttt{The de Sitter conjecture}
\item\texttt{The weak gravity conjecture}
\item\texttt{The trans-Planckian censorship conjecture}
\item\texttt{The scalar weak gravity conjecture}
\item\texttt{The bound on entropy conjecture}
\item\texttt{The no global symmetries conjecture}
\item\texttt{The no parametrically large global symmetries conjecture}
\item\texttt{The sublattice conjecture}
\item\texttt{The gradient conjecture}
\item\texttt{The dS distance conjecture}
\item\texttt{The trans-Planckian censorship principle}
\item\texttt{The dS swampland conjecture}
\item\texttt{The swampland dS conjecture}
\item\texttt{The dS conjecture}
\item\texttt{The averaged null energy condition conjecture}
\item\texttt{The trans-Planckian censorship criterion}
\item\texttt{The quantum gravity conjecture}
\item\texttt{The string swampland conjecture}
\item\texttt{The swampland distance conjecture}
\item\texttt{The swampland de Sitter conjecture}
\item\texttt{The swampland weak gravity conjecture}
\item\texttt{The swampland trans-Planckian censorship conjecture}
\item\texttt{The no global symmetries in quantum gravity conjecture}
\item\texttt{The no parametrically large global symmetries in quantum gravity conjecture}
    \end{enumerate}
\end{displayquote}

In the previous paragraph, we already offered sources for numbers one, two, and four. Number three is also a widely discussed conjecture, see. e.g. \cite{arkani-hamed_string_2007,Heidenreich_2016,palti_swampland_2019,ooguri_new_2017,Harlow_2016,grana2021swampland,Palti_2020,vanbeest2021lectures,brennan_string_2018,Palti_2017,vanbeest2021lectures}. Number five also exists, see e.g. \cite{palti_swampland_2019, palti_weak_2017,gonzalo_strong_2019}. To our knowledge, there is no conjecture named \textit{the bound on entropy conjecture}. However, entropy bounds play an important role in connections between the species scale, the distance conjecture, the weak gravity conjecture, and the de Sitter conjecture \cite{palti_swampland_2019}. Number seven exists \cite{harlow2018symmetries,Harlow_2019,palti_swampland_2019,Banks:1988yz,brennan_string_2018,Banks:2010zn}. Number eight is made up. It sounds like a combination of the distance conjecture and the no global symmetries conjecture. Number nine exists and is a combination of the distance conjecture and the weak gravity conjecture \cite{Heidenreich_2016,Heidenreich_2017,Klaewer_2017}. Number ten is a conjecture in mathematics, but not (yet?) in the swampland programme. Numbers eleven through fifteen are merely combinations of the names of existing conjectures. Number sixteen was new to the author of this letter, but was somewhat discussed roughly a decade ago \cite{kontou_averaged_2015,kelly_holographic_2014,kontou_quantum_2015,visser_gravitational_1997}. The remaining conjectures on the list are just word baublery, i.e. combinations of existing terms but not original conjectures. This example shows two things: First, \texttt{ChatGPT} cannot separate a concept from a word. If several words describe the same concept, it is not necessarily able to identify the synonyms\,---\, it often can provide synonyms for words, however, in this example it states different names for the same conjecture as different conjectures. Second, \texttt{ChatGPT} presents a fabricated response very confidently. In this example, it tries to comply with the request, without indicating that there actually is no list of 25 conjectures and that it made up most of them. Nevertheless, it achieves impressive results in terms of naming, and it makes certain to incorporate well-known conjectures. Nevertheless, it is also possible the prompt was not well-written, since it did not specify that it has to be a list of 25 unique and independent conjectures.

To further examine \texttt{ChatGPT}'s capabilities, we focus on specific swampland conjectures and ask \texttt{ChatGPT} for definitions and relations in the following subsections.

\subsection{The Weak Gravity Conjecture}
Gravity is the weakest force. No observation was made so far to contradict this statement. According to the \gls{wgc}, this is true for all string theoretical compactifications \cite{arkani-hamed_string_2007}. More precisely, the \gls{wgc} says that there is a stable particle whose repulsive gauge force exceeds the attractive gravitational force and which minimises the mass-to-charge ratio $m/q$, i.e. $\abs{m/q}<1$ \cite{arkani-hamed_string_2007}.

A more formal definition of the \gls{wgc} reads as follows:
In a theory with a $U(1)$ gauge symmetry; gauge coupling $g$; and action
\begin{equation}
    S=\int\!\mathrm{d}^dX\sqrt{-g}\left[\left(M^d_p\right)^{d-2}\frac{R^d}{2}-\frac{F^2}{4g^2}+...\right],
\end{equation}
with $d$ the number of dimensions of the \gls{eft}, $R^d$ the $d$-dimensional Ricci scalar, $F^2=\frac{1}{q!}F_{\mu_1...\mu_q}F^{\mu_1...\mu_q}$ the strength of the gauge field $A_{\nu...}$ \cite{Heidenreich_2016}, e.g. with the relation $F_{\mu\nu}=\frac{1}{2}\partial_{[\mu}A_{\nu]}$, and $M_p^d$ the $d$-dimensional Planck mass; a particle with mass
\begin{equation}\label{eq:ewgc}
    m\leq\sqrt{\frac{d-2}{d-3}}gq\left(M^d_p\right)^{\frac{d-2}{2}},
\end{equation}
$q$ the charge of the particle, exists \cite{palti_swampland_2019,Heidenreich_2016}.

\texttt{ChatGPT} knows the following about the \gls{wgc}:
\begin{displayquote}
    \texttt{The weak gravity conjecture (WGC) is a conjecture in theoretical physics that proposes a connection between the strength of a theory's gravitational force and the existence of charged particles or objects with extremely large charge-to-mass ratios. The conjecture was originally proposed by Cumrun Vafa and Alexander Westphal in 2006, and it has since become an important concept in the study of string theory and the swampland program.}

    \texttt{The basic idea behind the WGC is that in any theory that contains a consistent low-energy effective description of gravity, there must be at least one particle or object with a charge-to-mass ratio that is larger than the charge-to-mass ratio of a extremal black hole in that theory. This means that in any theory that satisfies the WGC, it should be possible to create a particle or object with a large enough charge-to-mass ratio that it can overcome the gravitational force and escape to infinity.}

    \texttt{The WGC has been motivated by the desire to understand the behavior of charged particles and objects in theories with a consistent low-energy effective description of gravity, and it has led to the development of several important results and insights in the field of theoretical physics. However, the conjecture is still being actively studied and is not yet fully understood.}
\end{displayquote}

Nothing in this statement is obviously wrong, as it remains rather vague while naming the relevant concepts that one might expect being mentioned. However, the remark \enquote{\texttt{The conjecture was originally proposed by Cumrun Vafa and Alexander Westphal in 2006,\textelp{}}} is interesting. When prompted for a source, \texttt{ChatGPT} responds with
\begin{displayquote}
    \texttt{The weak gravity conjecture was originally proposed by Cumrun Vafa and Alexander Westphal in a paper published in 2008 titled \enquote{A Stringy Test of the Fate of the Landscape}. The paper was published in the journal Physical Review D and can be found online at the following link:} \url{https://journals.aps.org/prd/abstract/10.1103/PhysRevD.78.046003}
\end{displayquote}
\texttt{ChatGPT} even provides a summary of this paper.\footnote{\texttt{In this paper, Vafa and Westphal proposed the weak gravity conjecture as a way to address certain problems that arise in the context of string theory, including the existence of a vast landscape of possible vacua and the difficulty of constructing stable de Sitter vacua. The conjecture suggests that in any consistent theory of quantum gravity, there must exist particles or fields with charge-to-mass ratios that are larger than a certain critical value. This has implications for the stability of de~Sitter vacua and the nature of black hole entropy. The weak gravity conjecture has since become an important part of the swampland program, and it continues to be studied and refined by researchers in the field.}}
In its previous answer, \texttt{ChatGPT} mentioned a paper written in 2006. Here, it provides a paper written in 2008. It sticks to the authors, though. However, no such paper exists! The two authors never published together, and the title is fictitious. The provided link leads to the paper \textit{Toward an AdS/cold atoms correspondence: A geometric realization of the Schrödinger symmetry} by D. T. Son \cite{son_toward_2008}. A better answer might have been \citet{arkani-hamed_string_2007}. This again shows that the user of \texttt{ChatGPT} has to be cautious when using \texttt{ChatGPT}'s answers. They are grammatically correct but not necessarily factual.

When asked for the most important, peer-reviewed, and published papers about the weak gravity conjecture, \texttt{ChatGPT} returns the following:
\begin{displayquote}
    \begin{itemize}
        \item \texttt{"The Weak Gravity Conjecture" by N. Arkani-Hamed, L. Motl, A. Nicolis, and C. Vafa (Journal of High Energy Physics, 2006). This paper introduced the concept of the weak gravity conjecture and discussed its implications for quantum gravity.}
        \item \texttt{"The Weak Gravity Conjecture from Emergence" by J. Kim, B. Vercnocke, and T. Wiseman (Journal of High Energy Physics, 2017). This paper discussed the relationship between the weak gravity conjecture and emergent gravity, and proposed a new version of the conjecture.}
        \item \texttt{"The Weak Gravity Conjecture and Emergence of Strong Coupling" by J. Kim, B. Vercnocke, and T. Wiseman (Physical Review Letters, 2018). This paper discussed the relationship between the weak gravity conjecture and the emergence of strong coupling, and proposed a new version of the conjecture.}
        \item \texttt{"The Weak Gravity Conjecture and Swampland Conjectures" by N. Arkani-Hamed, Y. Bai, and R. Sundrum (Journal of High Energy Physics, 2020). This paper discussed the relationship between the weak gravity conjecture and other swampland conjectures, and proposed a new version of the conjecture that incorporates these other conjectures.}
    \end{itemize}
\end{displayquote}

All the papers it provides have eerily similar titles\,---\,and not a single one does actually exist. The titles are intriguing. Even the brief summaries sound plausible. Some of the authors even publish in this or similar branches of physics (N. Arkani-Hamed, L. Motl, A. Nicolis, C. Vafa \cite{arkani-hamed_string_2007}, B. Vercnocke \cite{kallosh_analytic_2014,kallosh_string_2016}, R. Sundrum \cite{inspire_raman_2023}) while others publish in other fields or are made up (J. Kim, T. Wiseman, Y. Bai).

The responses it provides sound convincing and rich in detail, even though everything is wholly fabricated. This might make its answer unreliable, but can, if wisely done,  used to our advantage! Specifically, it can be used to find interesting and helpful analogies. For instance, we might be looking for a more down to earth analogy and ask \texttt{ChatGPT} to explain the \gls{wgc} in the words of a farmer:
\begin{displayquote}
    \texttt{The weak gravity conjecture is a concept in theoretical physics that proposes that the strength of a theory's gravitational force is related to the existence of charged particles or objects with extremely large charge-to-mass ratios. Essentially, this means that the stronger the gravitational force, the more likely it is that there are charged particles or objects present that can affect the force.}

    \texttt{From a farmer's perspective, this concept could be compared to the importance of understanding and managing the weather and climate conditions in a particular region. By understanding how different weather patterns and phenomena (such as storms, wind, and humidity) can affect their crops, farmers can make informed decisions about when to plant, water, and harvest, and can optimize the productivity and sustainability of their fields.}

    \texttt{Similarly, the weak gravity conjecture suggests that the strength of a theory's gravitational force can be analyzed and predicted based on the presence of charged particles or objects. This concept can be useful for farmers who are looking to optimize the productivity and sustainability of their fields, by helping them to understand and manage the various forces and phenomena that can affect their crops.}
\end{displayquote}

It's a bit far-fetched to assume that the \gls{wgc} can be used to optimise agricultural productivity. However, to communicate the concept in outreach or educational activities, drawing parallels between the environmental conditions' impact on the yield and the impact of the description of the gravitational force on the observed particles, might be helpful.

\begin{figure}[htbp]
\centering
\includegraphics[width=\linewidth]{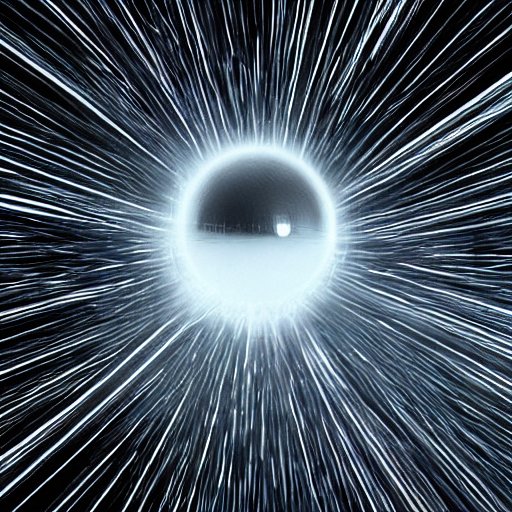}
\caption[Weak Gravity Conjecture]{\texttt{ChatGPT} description of an image representing the weak gravity conjecture: \enquote{\texttt{The image should depict a bright, glowing orb representing the strength of the gravitational force, surrounded by swirling darkness representing charged particles or objects affected by this force. The overall tone of the image should convey a sense of wonder and excitement, as the viewer approaches the orb and learns more about this mysterious and elusive concept in theoretical physics.}} Image created by \textit{Stable Diffusion} \cite{stable_diffusion_online_stable_2023}.}\label{f:WGC}
\end{figure}

Especially in outreach and education, we like to work with images and representations of concepts. We can use \texttt{ChatGPT} to generate visual representations of a concept's core ideas. We asked \texttt{ChatGPT} to describe how it envisions the weak gravity conjecture. The response served as a prompt for \textit{Stable Diffusion}\footnote{\url{https://stablediffusionweb.com}}, which then creates an image based on \texttt{ChatGPT}'s description. The outcome is shown in \cref{f:WGC}. Admittedly, this is more of an aesthetic picture than a simplified illustration of the concept.

\subsection{de Sitter Conjecture}
Another widely discussed swampland conjecture is the de~Sitter conjecture. It states that for a scalar field potential $V(\phi)$
\begin{equation}\label{eq:dSc}
\abs{\grad V}\geq \frac{c}{M_P}\cdot V,
\end{equation}
with $M_P$ the Planck mass and $c$ a constant of order one \cite{palti_swampland_2019,Denef_2018,Ooguri_2019,obied2018sitter}. It is a strong conjecture, as it not only rules out metastable de~Sitter vacua but also local maxima and saddle points \cite{Denef_2018}. When the acceleration is driven by dark energy, then dark energy must be dynamical, i.e. the equation of state is time-dependent, if dark energy is represented by a rolling scalar field \cite{palti_swampland_2019,Agrawal_2018}.

When asked what the de~Sitter conjecture implies for the standard model of cosmology, \texttt{ChatGPT} replies:
\begin{displayquote}
    \texttt{The de Sitter conjecture is a conjecture in theoretical physics that states that it is impossible to construct a stable de~Sitter space (a type of space with a positive cosmological constant) within the framework of string theory. This conjecture has been motivated by the desire to understand the behavior of theories with a positive cosmological constant and the challenges that they pose to the standard model of particle physics.}
    
    \texttt{If the de~Sitter conjecture is true, it could have significant implications for the standard model of cosmology, which is the current leading theory of the origin, evolution, and ultimate fate of the universe. The standard model of cosmology is based on the idea that the universe is expanding at an accelerating rate, driven by a mysterious form of energy known as dark energy. The standard model of cosmology also includes the hypothesis that the universe underwent a period of rapid expansion known as cosmic inflation in the very early universe.}

    \texttt{If the de~Sitter conjecture is true, it would suggest that the standard model of cosmology may need to be revised in order to be consistent with string theory. This could involve modifying or replacing the current hypothesis of cosmic inflation or finding an alternative explanation for the observed accelerating expansion of the universe. The de~Sitter conjecture is an active area of research in the field of theoretical physics, and it is still not fully understood. Further research will be needed to determine the implications of the de Sitter conjecture for the standard model of cosmology and other areas of physics.}
\end{displayquote}

\texttt{ChatGPT} is able to provide a relatively vague, surface-level explanation of the key concepts. It identifies accelerated expansion and inflation as the most important aspects of the concordance model that may need to be modified. \texttt{ChatGPT} learns from the written word. In the context of the de~Sitter conjecture, the most often discussed concepts are dark energy, accelerated expansion, and inflation. Therefore, \texttt{ChatGPT} can make this connection. However, it cannot genuinely connect concepts and infer implications no one else has yet thought of. This is also illustrated by the following example.

The relationship between the de~Sitter conjecture and the weak gravity conjecture is rarely studied. \texttt{ChatGPT} describes the relation between the de~Sitter conjecture and the weak gravity conjecture as follows:
\begin{displayquote}
    \texttt{The de Sitter conjecture and the weak gravity conjecture (WGC) are two separate conjectures in theoretical physics that have been proposed as part of the swampland program in string theory. However, there is a connection between these two conjectures, as they both involve the concept of a positive cosmological constant and the behavior of charged particles or objects in theories with a consistent low-energy effective description of gravity.}
    
    \texttt{The de Sitter conjecture states that it is impossible to construct a stable de Sitter space (a type of space with a positive cosmological constant) within the framework of string theory. This conjecture has been motivated by the desire to understand the behavior of theories with a positive cosmological constant and the challenges that they pose to the standard model of particle physics.}

    \texttt{The WGC, on the other hand, proposes a connection between the strength of a theory's gravitational force and the existence of charged particles or objects with extremely large charge-to-mass ratios. The WGC has been motivated by the desire to understand the behavior of charged particles and objects in theories with a consistent low-energy effective description of gravity.}
    
    \texttt{Although the de Sitter conjecture and the WGC are two separate conjectures, they are related in the sense that they both involve the concept of a positive cosmological constant and the behavior of charged particles or objects in theories with a consistent low-energy effective description of gravity. They are both active areas of research in the field of theoretical physics and have the potential to provide important insights into the fundamental nature of the universe and the structure of string theory.}
\end{displayquote}

Since there is no mention of gravity in the de~Sitter conjecture and no mention of the cosmological constant in the weak gravity conjecture, the response is largely incorrect. Admittedly, it is challenging to find analogies and connections between the two conjectures. In the literature, one frequently encounters links between the weak gravity conjecture and the distance conjecture \cite{Palti_2017,palti_swampland_2019,Grimm_2018,Heidenreich_2017,grana2021swampland,Heidenreich_2019} and between the distance conjecture and the de~Sitter conjecture \cite{Storm_2020,barrau2021string}. It is expected that some analogies could be drawn. However, no direct relations are known and \texttt{ChatGPT} is at this stage not able to identify them.

We find that \texttt{ChatGPT} does not create truly novel knowledge. Nonetheless, it can assist us to think about concepts in new terms. We've shown that \texttt{ChatGPT} can draw some elusive parallels between agriculture and the weak gravity conjecture. Here, we provide another example, by asking \texttt{ChatGPT} to explain the de~Sitter conjecture in terms of a financial advisor:
\begin{displayquote}
    \texttt{The de Sitter conjecture is a concept in theoretical physics that relates to the stability of certain spaces known as de Sitter spaces. These spaces are characterized by a positive cosmological constant, which means that they are expanding at an accelerating rate.}

    \texttt{In the framework of string theory, de Sitter spaces are believed to be unstable, which has important implications for our understanding of the universe and the fundamental nature of reality. Essentially, it means that certain types of spaces and configurations may not be as stable or long-lasting as we thought, and that we need to be cautious when considering them as possible models or explanations for the universe.}

    \texttt{From a financial perspective, this could be compared to investing in a company or asset that seems stable and promising at first, but ultimately turns out to be unstable or risky. It's important to do thorough research and analysis before making any investments, and to be aware of the potential risks and uncertainties that may be involved. In the same way, it's important to carefully consider the implications of the de Sitter conjecture and to approach it with caution and skepticism.}
\end{displayquote}

Both, financial markets and de~Sitter space require thorough investigation to identify instability points and can be in configurations that are (quasi)stable over a limited period. To be cautious is never bad advice, neither when it comes to investment decisions nor when it comes to model universes and theory development.

\begin{figure}[htbp]
\centering
\includegraphics[width=\linewidth]{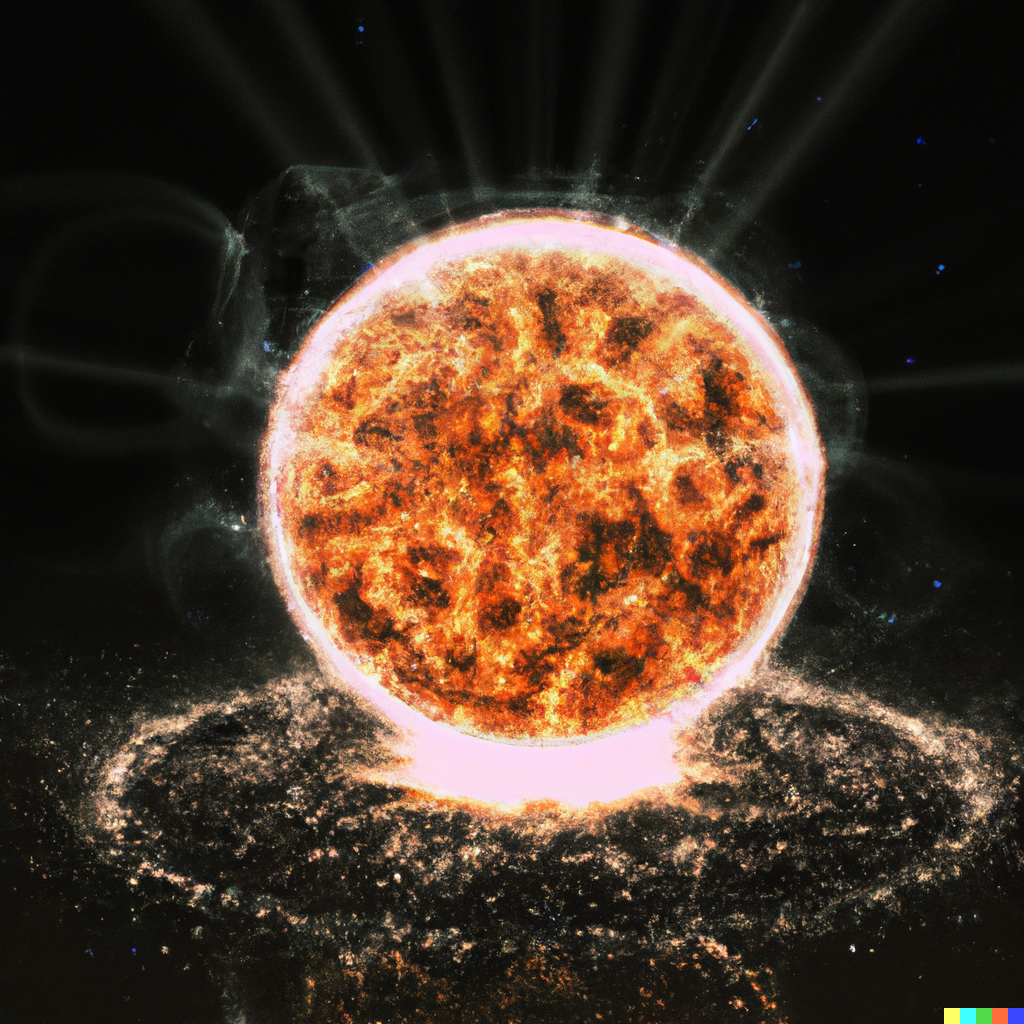}
\caption[de Sitter Conjecture]{\texttt{ChatGPT}'s description of an image representing the de~Sitter conjecture: \enquote{\texttt{The image should depict a bright, glowing sphere representing a de Sitter space, surrounded by a vast and complex landscape of theories and configurations. The sphere should represent the stability of the de Sitter space, and the landscape around it should represent the theories and configurations that are affected by this stability. The overall tone of the image should convey a sense of wonder and excitement, as the viewer approaches the sphere and learns more about this mysterious and elusive concept in theoretical physics.}} Image created by \textit{DALL-E 2} \cite{openai_dalle_2023}.}\label{f:dSC}
\end{figure}

We conclude this section by \cref{f:dSC}\footnote{\Cref{f:dSC} was created by \textit{DALL-E 2}: \url{https://openai.com/dall-e-2/}.}, which represents the de~Sitter conjecture and could be used as a cover image for a book chapter or a slide deck.

\section{Outlook}
In the previous paragraphs, we show that \texttt{ChatGPT} can synthesise various concepts and present them coherently. It succeeds in highlighting the core ideas and identifying the relevant keywords. It excels at combining what is known. However, it cannot genuinely make new connections or create new knowledge. Nevertheless, it is a useful tool for education and outreach. It helps us to rephrase and find the right words for our target audience. It can be used to formulate analogies and find parallels to other fields. It can summarise what we already know. In the module \textit{Physics in Society}\footnote{A module on the history and philosophy of physics and science.} at Durham University, \texttt{ChatGPT} scored \SI{71\pm2}{\percent}, where students score \SI{71\pm5}{\percent} on short, 300-words essays \cite{yeadon_death_2022}. The same case study shows low, single-digit percentages for plagiarism using \textit{TurnitIn}\footnote{\url{https://www.turnitin.com}} and \textit{Grammarly}\footnote{\url{https://www.grammarly.com/plagiarism-checker}}. It is worth noting that there are tools to indicate the use of GPT\footnote{Under \url{https://openai-openai-detector.hf.space}, an online tool can be found, which uses the GPT 2 dataset \cite{openai_gpt-2-output-datasetdetector_2021,noauthor_gpt-2_nodate}}. However, detection could be evaded by using rephrasing tools such as \textit{QuillBot}\footnote{\url{https://quillbot.com}}. It is up to the teacher to decide whether essays are still a meaningful form of assessment. \texttt{ChatGPT} is a new and powerful tool to craft short essays. It can help non-native speakers overcome language barriers by assisting them in formulating their ideas and insights.

Posters are another application, where \texttt{ChatGPT} comes in handy. Assume you want to make a poster for a general science venue to present the swampland conjectures. You can then ask \texttt{ChatGPT} to write short paragraphs on a selection of swampland conjectures, a brief introduction to string theory and the swampland programme, prompt it to suggest a catchy title, and even have it describe some illustrations, which you can then create using other AI tools like \textit{Stable Diffusion}, \textit{DALL-E}, or \textit{MidJourney}\footnote{\url{https://midjourney.com}} (see \cref{f:WGC,f:dSC} for examples). Then, you include the relevant formulas you want to discuss, create graphs that adhere to scientific standards, and weave everything together.  You can concentrate on the poster's conceptualisation because you spend less time writing the paragraphs.

In the near future, we will see a variety of AI-tools. They are already powerful companions that can be used in tandem. A \texttt{ChatGPT} produced text can be rephrased using \textit{QuillBot}. An image description by \texttt{ChatGPT} can be used as an input prompt for \textit{DALL-E}, \textit{Stable Diffusion}, or \textit{MidJourney} to create a visualisation. \texttt{ChatGPT} can be used to write a whole script for a video tutorial, which can then be used in \textit{Fliki}\footnote{\url{https://fliki.ai}}, to create a video, based on the script, including animations and a computer-generated voice. These tools are already very user-friendly, but further development is underway. Moreover, with GPT-4, the framework behind ChatGPT will get a major upgrade in the near future \cite{romero_gpt-4_2022}.

\section{Conclusion}
As a natural language processing model, \texttt{ChatGPT} is naturally good with words. If assigned a task it can only partially fulfil, it uses its abilities to come up with an answer and presents it with full confidence, even if the answer is inaccurate. This can be used to one's advantage, if one is able to identify the made up parts of an answer. Perhaps those parts start a thought process that leads to new insights. Asking \texttt{ChatGPT} to respond in the style of a given persona can help us find new and helpful parallels to other fields. This is of great use for educational and outreach endeavours. \texttt{ChatGPT} is an always-on sparring partner who is eager to assist and collaborate, when it comes to testing ideas and concepts. At its current state, it is not well suited to gain new knowledge or answer genuine questions. It will respond with absolute assurance, but the answer might be utterly false and completely fabricated. Nevertheless, its capabilities are astounding. We are on the brink of an AI-aided era of advancement if used with caution.

\section{Disclaimer}
The creation of this letter was assisted by the \textit{chatbot} \texttt{ChatGPT} by OpenAI \cite{openai_chatgpt_2022}. Specifically, everything written in typewriter is a \texttt{ChatGPT} output. Furthermore, the title of this letter was proposed by \texttt{ChatGPT} itself. The images were created by \textit{DALL-E 2} \cite{openai_dalle_2023} and \textit{Stable Diffusion} \cite{stable_diffusion_online_stable_2023}. 

Kay Lehnert is a recipient of the John and Pat Hume Scholarship and acknowledges support from the Swiss Study Foundation.

The average power consumption per request of \texttt{ChatGPT}, \textit{DALL-E} or \textit{Stable Diffusion} are not public. Furthermore, it is not clear whether the energy used to train the models should also be taken into account, when estimating the CO$_2$-equivalent of this article. The author's personal computer caused less than \SI{100}{g} CO$_2$e and is compensated by \url{https://climeworks.com}. To internalise possible additional external factors, \SI{1}{kg} of CO$_2$ was compensated to cover for this work. Therefore, we consider this work as \textit{probably} carbon-neutral.
\FloatBarrier